# Failure Detection and Recovery in Hierarchical Network Using FTN Approach

Bhagvan Krishna Gupta[1], Ankit Mundra[2], Nitin Rakesh[3]

[1, 2, 3]Computer Science and Engineering Department, Jaypee University of Information Technology, Waknaghat, Solan, Himachal Pradesh 173234, India

**Abstract**

In current scenario several commercial and social organizations are using computer networks for their business and management purposes. In order to meet the business requirements networks are also grow. The growth of network also promotes the handling capability of large networks because it counter raises the possibilities of various faults in the network. A fault in network degrades its performance by affecting parameters like *throughput*, *delay*, *latency*, *reliability* etc. In hierarchical network models any possibility of fault may collapse entire network. If a fault occurrence disables a device in hierarchical network then it may distresses all the devices underneath. Thus it affects entire networks performance. In this paper we propose *Fault Tolerable hierarchical Network* (FTN) approach as a solution to the problems of hierarchical networks. The proposed approach firstly detects possibilities of fault in the network and accordingly provides specific recovery mechanism. We have evaluated the performance of FTN approach in terms of delay and throughput of network.

## 1. Introduction

Internet technology has provided valuable means of communication network model to its users. For the corporate network design process to meet organizations business and technical requirements, it is very necessary to provide a network topology. Several network topologies have been introduced depending on their need for the communication process. Whereas in the business organizations, it is proficient to use 'divide and conquer' approach for design a network. This approach develops the network design in layers. These layers design corresponds to hierarchical network architecture. In this architecture each layer has some specific functions. For example, a layer contain high-speed routers that carry traffic across the enterprise sections, another layer contains medium-speed routers that connect buildings at each campus of the enterprise [1]. On the other hand in hierarchal networks it is very challenging task to provide reliable communication over internet due to possibility of faults in networks [2]. These faults can make the networking device as a dumb terminal and it stopped working until the fault is repaired. And this dumb terminal also affects the entire next level node connected to it in the form of hierarchical architecture. In hierarchical networks various faults can occurs i.e. physical defects, hardware malfunction, link corruption (cable damage), IP connectivity errors, physical change in topology, network misconfiguration, and electrical noise [3].

In this paper we discussed two main issues seeing in hierarchical networks during any types of fault happening. We consider a scenario, which shows if any fault occurs in a network device then it will tend all the network devices (underneath to it) become either inactive for a specific time or permanently stopped working. First issue is fault detection and second is fault recovery [4]. Fault detection techniques work in the context to alert other network devices regarding the faulty device. Further network administrator is responsible for fault recovery. Generally many faults are repaired manually and some specific types of faults are repaired using software applications for e.g. network malfunctioning fault.

In the conventional approaches of fault detection and recovery in networks, they suggest retransmission of message. Means the sender device has to retransmit the message if it does not receive acknowledgment before a specific time (due to any fault in the network). But in the situation where sender retransmits the message repeatedly for number of times then it results in increase delay, latency, reduce throughput, wastage of bandwidth of network and also create congestion in network. Further fault recovery is used to manage the network device, traffic and provides reliable communication. To overcome this problem we propose FTN approach for fault detection and fault recovery. FTN approach provides algebraic formulation of different metrics for e.g. traffic distributed over the network, total packet loss and expected buffer size of the network. FTN algorithm works on three phases first phase for calculating buffer size of router, second phase deals with fault detection whereas third phase deals with fault recovery.

In FTN approach we use basic message format (Fig. 1) for deriving algorithm of fault detection and fault recovery. This message format contains four fields i.e. flag, sender address, destination address, data part [5-6]. Flag is 1 byte field where as sender address and destination address is 4 byte field. Data part may contain up to 1500 byte.

| Flag | Sender address | Destination address | Data (optional) |
|------|----------------|---------------------|-----------------|

Fig.1 Message Format

This paper consist six sections. First section briefly introduces the hierarchical network architecture and various possible faults in that architecture along with proposed approach. Section second illustrates the related work for fault detection and recovery process. Further section third describes the problem formulation by considering a scenario with two failure cases i.e. device failure and link failure. Section fourth describes the proposed approach along with the algorithm and algebraic formulation. Section fifth illustrates the performance evolution of proposed approach and comparison between proposed approach and conventional approaches of fault detection. Finally we conclude in section sixth.

## 2. Related Work

This section illustrates the various approaches proposed for detecting fault in hierarchical networks. The earlier approach was sending ping connectivity message to other adjacent device. If ping reply comes from other device then it shows that other device active. It means a path is available between ping requestor and the other device. In this approach sender manually initiate the process of sending ping message to other devices [5-7]. The other approach of fault detection is suggesting use of using routing protocol. In this each router periodically transmit routing table to all adjacent devices. [5, 8] proposed a routing table which maintains status of all the paths between network devices. Path with inactive status shows that the device is faulty.

Saurabh *et al.* proposes a hierarchical framework for providing fault tolerance in the hierarchical networks. It introduces the software implemented fault tolerance layer of a distributed environment [9].

Later on Heman Pathak *et al.* propose solution for dynamic grid environment failures (e.g. Link down, Resource failure). They present a fault tolerance scheme for Hierarchical Dynamic Scheduler (HDS) for grid workflow applications [10].

## 3. Problem Formulation

For formulating the problem of fault detection and recovery in the hierarchical networks we consider hierarchical network architecture of any corporate organization (Fig. 2). This architecture contains three types of devices one is a Group Server1 (GS1) to represent organization head office. Secondly seven routers name as R1, R2, R3, R4, R5, R6, and R7. And five switches (SW1, SW2, SW3, SW4, SW5) to provide connectivity to multiple end devices. Further GS1 is connected to the router R1 for next level communication. We consider, routers are connected to group server, router, switch or member host using bidirectional link. Now, when GS1 sends message to other device for e.g. SW3-1 (here -1 represent first system connected to SW3). Then data will forward using path

$$GS1 \rightarrow R1 \rightarrow R3 \rightarrow R6 \rightarrow SW3 \rightarrow 1$$

If end device 2 connected to switch 4 (SW4) interested to transmit the message to device 3 of switch 2(SW2) then forwarding of message will follow the path:

$$2 \rightarrow SW4 \rightarrow R3 \rightarrow R1 \rightarrow GS1 \rightarrow R2 \rightarrow R5 \rightarrow SW2 \rightarrow 2$$

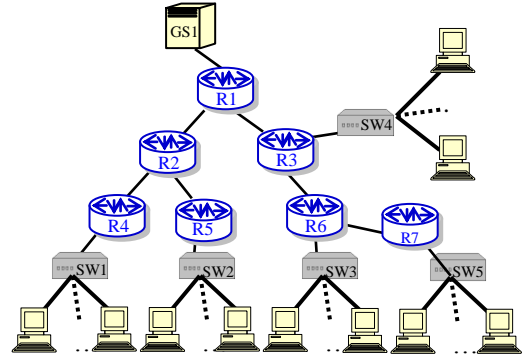

Fig. 2 Hierarchical Model.

Now, there are two main problems associated with hierarchical model, which are describe below:-

3.1 Router or Networking device failure

In hierarchical model (Fig. 3) if any routing device becomes faulty then it stops forwarding of message to next level devices. For e.g. consider router R3 failed due to any of above explained reason (section 1) then all the traffic forwarded by R3 will stopped. In this case due to failure of R3 the adjacent devices are also stopped working i.e. SW4, R6 (R7, SW3, SW5). When GS1 wants to send the message to the host connected with switch SW5 then the message does not deliver because of inactive path (R3-R6-R7-SW5)

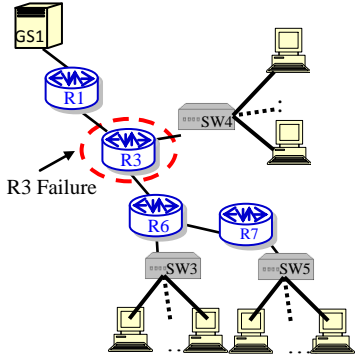

Fig.3 Router fail

And forwarding of messages will remain discontinued until R3 get repaired. Therefore the entire messages which are forwarded via R3 are dropped and GS1 retransmitted the message after timeout.

### 3.2 Transmitting links failure

In hierarchical model (shown in Fig. 2), if any communication link becomes fail due to any fault then devices which are connected to this link will not communicate to each other. Thus messages that were forwarded between those devices are dropped. For e.g. in Fig. 2 links R1 to R3 failed then all the traffic towards R1 or R3 stopped until link is repaired. In this case again GS1 retransmitted the message after timeout.

## 4. FTN Approach

To overcome the problems discussed in previous section we propose an approach in order to provide reliable transmission with low latency, network consumption in hierarchical networks. This proposed approach is known as Fault Tolerable hierarchical Network (FTN). FTN works on three phases (calculation of buffer size; fault detection; and fault recovery).
First phase is based on calculating buffer size for storing message in intermediate router. For calculating buffer size we consider initially a scenario over a time period where numbers of devices are faulty. In this we store the message in router buffer if any routing device towards destination path is faulty.
Second phase is based on fault detection and second phase is fault recovery. In fault detection phase each router checks its adjacent devices, if any device identified faulty then it notify its parent device. Parent device in hierarchical network is the device resides at just above level of that node.
Third phase is the fault recovery, used to provide recovery mechanism from fault. During the FTN approach of fault detection and recovery we also need routing table of the routers. For this, here we take sample routing table of router R2 (Table 1). So as shown in the table routing table contains following fields i.e. Network Address, Next Hop, Interface, Connection Type, and Connection Status. Network Address field defines the IP address of the devices in the network. Next hop field is used to represents next node towards destination node. Interface shows the router interface from which data will forward. Connection type filed is used to show the type of connection either directs (represented by D) or indirect connection (represent by I). Connection status shows whether connection is active or not. If connection status value is 1 (one) means connection is active and 0 (Zero) means connection is inactive.

Table 1: Routing table of Router R2

| Network Address | Next Hop | Interface | Connection Type | Connection Status |
|---|---|---|---|---|
| 181.1.1.2 | – | 1 | D | 1 |
| 171.1.2.1 | – | 4 | D | 1 |
| 168.1.1.1 | – | 2 | D | 1 |
| 172.1.1.1 | 168.1.1.1 | 4 | I | 1 |
| 173.1.1.1 | 168.1.1.1 | 4 | I | 1 |
| 162.1.1.1 | 168.1.1.1 | 4 | I | 1 |
| 165.1.1.1 | – | 3 | D | 1 |

### 4.1 Algebraic formulation of buffer size calculation

Total traffic distributed over the network is calculated using probability distribution. In our approach Poisson distribution is used for finding out the total traffic flows in the network in a given time interval. Poisson distribution [12] (equation 1) is based on random distribution, it calculate random traffic distribute over single device.

$$X_N = \frac{\lambda t\ ^n e^{-\lambda t}}{n!} \quad …(1)$$

Now, for calculating total number of packet distribution over the network, means packet distribution over all networking device available is calculated by equation 2:

$$D = \sum_{1}^{N} (X_N = \frac{\lambda t\ ^n e^{-\lambda t}}{n!}) \quad …(2)$$

Where, $X$ represents Poisson distribution packet flow in network; $e$ is the base of the natural logarithm (i.e., e = 2.71828...); $\lambda$ is mean arrival rate; $t$ is time duration; n is variable; $N$ is total number of devices in the network.

Further when any node fails then it stops forwarding of data, hence data do not reach to their destination. After timeout that packet will drop, it results into packet loss.
Total packet loss on that device will depends on probability distribution function and duration of time device gets faulty, Number of packet loss can be calculated as:

For single device: $L = X_N \times T$ ... 3
For multiple devices: $L = X_N \times T \times K$ ... 4
As,
$$X_N = \frac{\lambda t^n e^{-\lambda t}}{n!} \quad so, \quad L = \frac{\lambda t^n e^{-\lambda t}}{n!} \times T \quad ...(5)$$

Here, $T$=Time interval device gets faulty, $K$= number of device gets faulty for that time period. If in hierarchical network packet loss if any route in invalid in routing table by using our approach we store message in routers buffer.

Now, we calculate expected size of router buffer as:
$$B = Y \times L \times packet\ size\ in\ bits \quad ...(6)$$
Here, B=Expected buffer size, Y= constant factor

## 4.2 Fault Detection algorithm

In fault detection algorithm, it uses Fault Detection Message (FDM) to detect fault in the network. FDM is send by the router devices to check their adjacent whether they are active or not. FDM is further classified in two types first is Fault Detection Query Message (FDQM) and second is Fault Detection Report Message (FDRM). FDM works similar as ping message because it is used for checking connectivity status between devices. The difference between ping message and FDM message is that it automatically checks connectivity between directly connected devices over fixed interval of time. FDM sends QM (Query message) or RM (Report message) to all the directly connected nodes. Then if any device receives FDQM then it replies to sending device by FDRM and if any device receives FDRM message then it does not need to send any message back to sender.

Table 2: Algorithm for Fault Detection

| Input:-Message M |
|---|
| Fault_Detection { <br> 1. Periodically send QM to all directly connected interface <br> 2. R = receive (M); <br> 3. J= first (flag);     \*check first bit of flag*/ <br> 4. S = Sender (M); <br> D = Destination (M); <br> P = Previous Hop (S); <br> N = Next Hop(D); <br> /* find route details of message using routing table*/ <br> 5. If (J = 0) <br>    then FDM; <br> Else <br>       data message goto step 8; <br> 6.  Q = second (flag) <br> /* check second bit of flag */ <br> 7. If (Q=0) <br>    then send[RM]→P <br> /*FDQM and send RM to sender or previous hop*/ <br> Else <br> FDRM <br> 8. C = Active ( P & R ) <br> /*active function is used to update routing table and make connection active in routing table */ <br> 9. U= Map ( D ) <br> /*Map function is used to check destination address is Unicast, Multicast, Broadcast*/ <br> 10. I=find(D) <br> \*find function used to find all destination interface */ <br> 11. y = 0 <br> 12. for (I =F;I≤L ; I=next location in I) <br> { <br> if(valid(I)) <br>    send[M] <br> else <br> y = y+1 <br> 13. if (y!=0) <br>    fault Recovery() <br> /* Call function fault recovery */ <br> 14. end of algorithm |
| *Note: R=Current routing device,* <br>       *y, C, Q, J = Variable* <br>       *I= Used for storing destination address* <br>       *F=first address in destination list.* <br>       *L=last address of destination list* |

## 4.3 Fault Recovery algorithm

Fault recovery algorithm is used to provide reliability in network and reduce latency of network by faster recovery from faults. In hierarchical network if any device detected faulty, then fault recovery algorithm is responsible for following tasks: 1. Stores the message at the buffer of previous router (preceding router to the faulty device); 2.Transmit the message after node get repaired; 3. If faulty device not repaired before timeout then it notifies the sending (source) device. Table 3 shows the Fault Recovery Algorithm:

Table 3: Algorithm for Fault Recovery

| Input: Size of Buffer=B <br> Time out of message at current device = $t_o$ <br> Periodic time for send QM to fault device= $t_p$ |
|---|
| Fault_Recovery { <br> 1. initially $t_o$=0 <br> 2. if( $M_j$<= B )\*check for messagesize not exceedsbuffer size* <br> { <br> B = store[$M_j$] <br> B= B-$M_j$ <br> \*store function issue to store message in buffer */ <br> } <br> 3. If( ! valid ( R & N ) ) <br>    For(k = 0; k < $t_o$; k = k+$t_p$) <br> { |

```
    Send[QM]→N
\*periodically send query message to faulty next hop*/
If( valid ( R & N ) )
\*check connection status is active or not*/
{
Send[M]→N;
\*send message to next hop*/
Active(R & N);
break;
}
}
4. If ( k > t₀ )
     Send[nack]→S
     \* timeout expires then send NACK to sender*/
5. B=clear(M→B)
\* clear message from memory buffer*/
6. B=B+M
\* reinitialise the size of memory buffer*/
7. End of algorithm
Note: k=variable
```

### 4.4 Flow diagram of FTN approach

By using flow diagram we show step by step working of FTN approach.

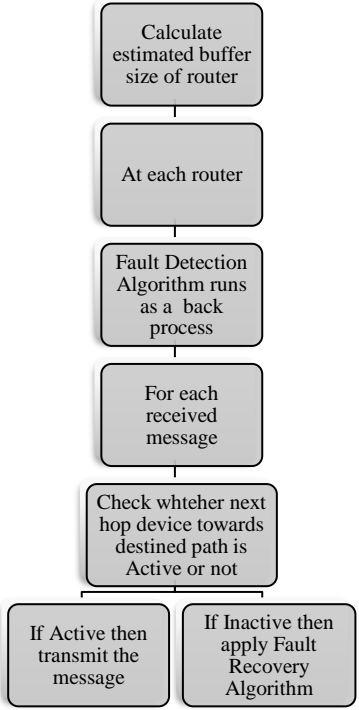

Fig. 4 Flow diagram of FTN approach

## 5. Performance Evaluation

For the performance evaluation of FTN approach in hierarchical network we consider two cases; first case is when there is no fault in the network and second is when network is faulty. For both the cases first we calculate buffer size required.

*Step 1: Calculating Buffer size:*

Here, we consider the scenario where network is faulty. To detect fault in the network FDQM is send by each router in every 500*ms*. We consider frame rate of 100frame/second and calculated all values in millisecond. Then, we first calculate total number of packet flow in network by assuming $t = 10, \lambda = 50, n = 100$.
$$X = 1.358 \times e^{-128} \quad \text{... from eq (1)}$$
In above architecture (Fig. 2) total 8 routing devices are present so the total packet distribution in the network is:
$$D = 1.086 \times e^{-127} \quad \text{... from eq (2)}$$
Now, to evaluate packet loss while $T = 20ms$, then one of the devices become faulty and total packet loss is:
$$L = 0.2716 \times e^{-127} \quad \text{... from eq (3)}$$
We consider that 4 devices became faulty at the same time period then total packet loss in network is:
$$L = 4 \times 0.2716 \times e^{-127}$$
$$= 1.0864 \times e^{-127} \quad \text{... from eq 4}$$
Here L represents total packet loses occurred over the device in per unit of time for e.g. second, millisecond.

Now for calculate buffer size of the router we consider total fault duration is 1000ms. So we divide this fault duration period into five parts as:
   0 to 200 ms only one device is faulty
   200 to 400 ms four devices faulty
   400 to 600 ms two device faulty
   600 to 800 ms three devices faulty
   800 to 1000 ms four devices faulty
Then, total packet loss during that period: $L = 2.716e^{-127} + 10.864e^{-127} + 5.432e^{-127} + 8.148e^{-127} + 10.864e^{-127}$
$$L = 38.024e^{-127}$$

Now, for calculating buffer size:
$$B = 38.024\ e^{-127} \times 10 \quad \text{... by using equation}(6)$$

*case1: Network is Fault Free:*

When there is no fault in the network then all the message will successfully reached to their destination. Forwarding of messages from intermediate devices will induce delay in the network i.e. transmission delay, propagation delay, queuing delay, processing delay. To evaluate performance of FTN we assume that only one device can forward and processing delay is negligible because each router is fast enough to process [13, 14].
Now, in Fig. 2 If GS1 sends data to device SW3-1 so total delay is
$$Delay = TD + SD + PD \quad \text{... (7)}$$

Where, TD is *Transmission Delay*, SD is *Switching Delay* and PD is *Propagation Delay*.
Then Latency and buffer clear timeout can be calculated as:
$$latency = 2 \times Delay \quad ...(8); And$$
$$Timeout\ for\ clear\ buffer = 2 \times latency \quad ...9$$

Further efficiency is calculated as below:
$$E = \frac{useful\ time}{total\ time} \quad ...(10)$$

Let us consider a scenario in which the network link capacity is 1Mbps and each frame contains 500 bits of data. Propagation delay between two adjacent nodes is 50ms. Furthermore we assume that GS1 send data at different frame rate to the receiving device is R7-SW2-1. Now, we calculate delay, latency, and throughput

Table 4: Delay and Efficiency Calculation when Router is Fault Free

| Frame rate | QD (s) | TD (s) | PD | Delay (s) | Latency (s) | Efficiency |
|---|---|---|---|---|---|---|
| 100 | 0 | 0 | 0.05×6 | 0.3 | 0.6 | 0.5 |
| 1000 | 0 | 0 | 0.05×6 | 0.3 | 0.6 | 0.5 |
| 2000 | 0 | (2000×500)/10$^6$ | 0 | 1.0 | 2.0 | 0.5 |
| 3000 | 3×0.5 | (3000×500)/10$^6$ | 0 | 3.0 | 6.0 | 0.25 |
| 5000 | 5×0.5 | (5000×500)/10$^6$ | 0 | 5.0 | 10 | 0.25 |
| 7500 | 7.5×0.5 | (7500×500)/10$^6$ | 0 | 7.0 | 14 | 0.26 |
| 10000 | 10×0.5 | (10000×500)/10$^6$ | 0 | 10 | 20 | 0.25 |

Table 4. shows the Delay, Latency, and Efficiency of network using equation 7, 8, 9, 10. Here 's' represents seconds.

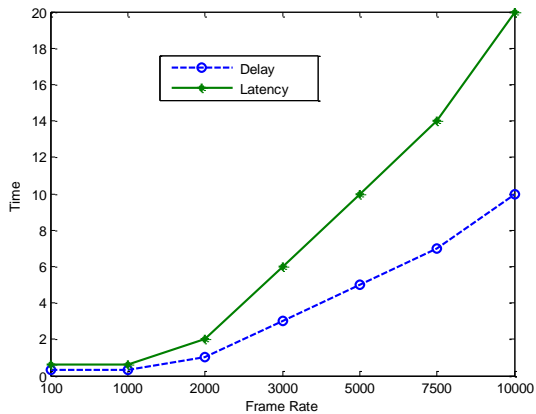

Fig. 4. Relation between Delay and Latency at different-different frame rate

Fig. 4 shows that both Delay and Latency get increase as the frame rate increases.
Now, we calculate throughput of the network over different frame rate:

$$data\ rate = frame\ rate \times number\ of\ bits\ in\ frame \quad ...(11)$$

$$throughput = \frac{data\ rate}{link\ capacity} \quad ...(12)$$

We consider link capacity is 1Mbps and each frame contains 500 bits.
Further Table 5 shows the throughput of the network calculated by equation 11.

Table 5: Throughput Table

| Frame rate | Data rate | Throughput |
|---|---|---|
| 100 | 100×500 | 0.05 |
| 500 | 500×500 | 0.25 |
| 1000 | 1000×500 | 0.50 |
| 1500 | 1500×500 | 0.75 |
| 2000 | 2000×500 | 1.00 |
| 2500 | 2500×500 | 0.75 |
| 3000 | 3000×500 | 0.50 |
| 3500 | 3500×500 | 0.25 |

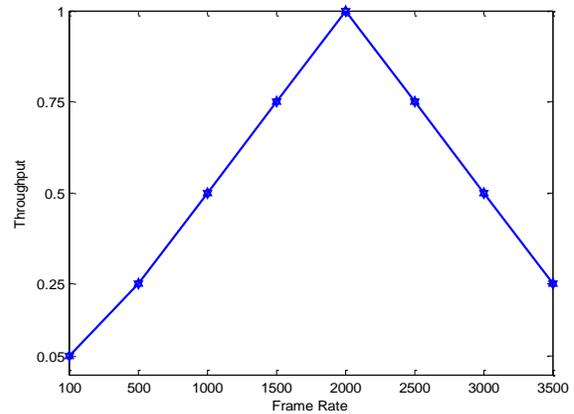

Fig.6. Throughput of network at different Frame Rate

Fig. 6 shows that the throughput of network gets increased till a specific frame rate reached after this it will decreases because some of frames are lost for exceeding link capacity.

*Case 2: Network is Faulty:*

When there is a fault present in the network then,
Here we calculate values of latency at frame rate= 100frame/second. Here we consider R3 is faulty so according to FTN approach:

*Step 1:* Each router will periodically send FDQM to all adjacent devices for e.g. R1 sends FDQM to R3 and R2. Consider periodic time for sending FDQM is 500ms.

*Step.2:* GS1 wishes to send data to R7-SW5-2 (device number 2 attached to switch 5).
*Step.3:* Consider R3 fails at time 0 then data transmitted by GS1 is reach router R1 at 50ms. In R1 it identifies R3 is faulty using routing table connection status value.
*Step.4:* R1 stores the message in buffer and start timeout for clear buffer memory for e.g. for SW7-2 timeout for clear buffer is 1000ms and if consider initial 50ms propagation time for GS1 to R1 then time out for clear buffer expires using FTN approach is at time 1050ms.
*Step.5:* Router R1 receives FDRM before timeout for clear buffer expires then it transmit the message to destination path for e.g. consider R3 is faulty duration 500ms then it receives FDRM before timeout expires so it transmit the message and clear the message from router buffer.
*Step.6(a):* Router R1 is not received FDRM message and timeout for clear buffer expired. then R1 clear the buffer message stored in R1 buffer destined towards router R3 and send negative acknowledgment to sender for e.g. if fault duration is 1500ms and timeout at router R1 is 1050ms so after timeout R1 clear the buffer and send negative acknowledgement to sender GS1
*Step.6(b):* When sending device receive negative acknowledgment from any routing device then it retransmit the message for e.g. here GS1 receives negative acknowledgment at time 1150ms from router R1 then GS1 again transmit the message.
*Step.6(c):* Message again reaches at router R1 check routing table connection status active or not for e.g. message reaches at 1250ms again wait for FDRM reply. It received FDRM at time 1550ms.
*Step.6(d):* R1 sends message to SW5-2 and SW5-2 receive message at 1800ms and acknowledgment will receive at GS1 on time 2100ms.
*Step.7*: Apply step 3 to 7 for different fault duration.

The values of delay and latency are calculated and shown in Table 6 using FTN approach.

Table 6: Latency comparison table when router is faulty

| Fault duration | Time out | Latency(Conventional Approach) | Time out | Latency (FTN Approach) |
|---|---|---|---|---|
| 0.500 | 1.200 | 1.800 | 1.050 | 1.100 |
| 1.000 | 1.200 | 1.800 | 1.050 | 1.600 |
| 1.500 | 2.400 | 3.000 | 2.150 | 2.100 |
| 2.000 | 2.400 | 3.000 | 2.150 | 2.600 |
| 2.500 | 3.600 | 4.200 | 3.250 | 3.100 |
| 3.000 | 3.600 | 4.200 | 3.250 | 3.600 |
| 4.000 | 4.800 | 5.400 | 4.350 | 4.100 |
| 4.500 | 4.800 | 5.400 | 5.450 | 5.100 |

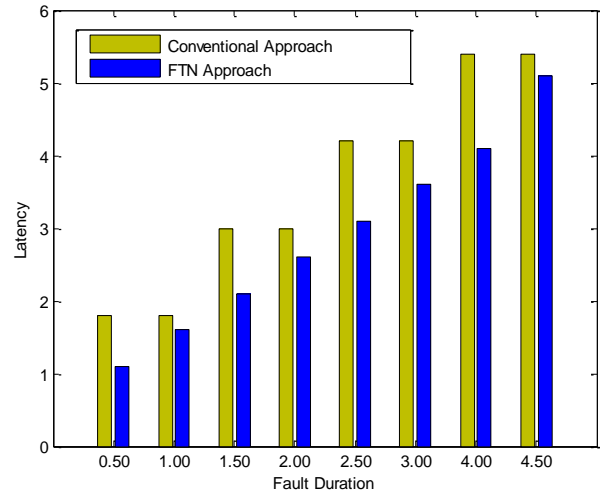

Fig. 7 Comparison between conventional approach and FTN approach

Fig. 7 shows the comparison based result between conventional approaches and FTN approach.

## 6. Conclusion

In this paper we have proposed an approach to handle fault in hierarchical networks. This approach basically deals with faults that interrupt the exchanging of information. In this paper we have evaluated the issues of network fault and have compared this with the conventional approach of retransmission of message. Our result shows that FTN approach is better than the conventional approach over the network parameters: *delay, throughput*. In future the FTN approach will be studied and extended for congestion in hierarchical network. Based on the results of congestions based FTN approach it will further be generalized for a variety of business networks.